\documentclass[fleqn,12pt,twoside]{article}
\usepackage{espcrc1}
\usepackage{epsfig}

\usepackage{graphicx}
\usepackage[figuresright]{rotating}

\newcommand{\AmS}{{\protect\the\textfont2
  A\kern-.1667em\lower.5ex\hbox{M}\kern-.125emS}}

\newcommand{\seff}{\s_{\rm eff}}
\newcommand{\s}{\sqrt{s}}
\newcommand{\pp}{pp}
\newcommand{\pbarp}{\overline{p}p}

\newcommand{\epem}{e^+e^-}

\newcommand{\nch}{N_{ch}}
\newcommand{\np}{N_{part}}

\newcommand{\ntot}{\langle\nch\rangle}
\newcommand{\avenp}{\langle\np\rangle}

\newcommand{\halfnp}{\langle\np/2\rangle}
\newcommand{\etap}{\eta^{\prime}}

\newcommand{\dndetap}{d\nch/d\etap}

\newcommand{\dndetapnp}{\dndetap / \halfnp}

\newcommand{\ratio}{\ntot/\halfnp}

\newcommand{\yb}{y_{\rm beam}}

\title{Recent Results from PHOBOS at RHIC}

\author{Robert Pak for the PHOBOS Collaboration \\ 
\mbox{} \\
\small{
B.B.Back$^1$,
M.D.Baker$^2$,
D.S.Barton$^2$,
R.R.Betts$^6$,
M.Ballintijn$^4$,
A.A.Bickley$^7$,
R.Bindel$^7$,
A.Budzanowski$^3$,
W.Busza$^4$,
A.Carroll$^2$,
M.P.Decowski$^4$,
E.Garc\'{\i}a$^6$,
N.George$^{1,2}$,
K.Gulbrand-\
sen$^4$,
S.Gushue$^2$,
C.Halliwell$^6$,
J.Hamblen$^8$,
G.A.Heintzelman$^2$,
C.Henderson$^4$,
D.J.Hofman$^6$,
R.S.Hollis$^6$,
R.Ho\l y\'{n}ski$^3$,
B.Holzman$^2$,
A.Iordanova$^6$,
E.Johnson$^8$,
J.L.Kane$^4$,
J.Katzy$^{4,6}$,
N.Khan$^8$,
W.Kucewicz$^6$,
P.Kulinich$^4$,
C.M.Kuo$^5$,
W.T.Lin$^5$,
S.Manly$^8$,
D.McLeod$^6$,
J.Micha-\
\l owski$^3$,
A.C.Mignerey$^7$,
R.Nouicer$^6$,
A.Olszewski$^3$,
R.Pak$^2$,
I.C.Park$^8$,
H.Pernegger$^4$,
C.Reed$^4$,
L.P.Remsberg$^2$,
M.Reuter$^6$,
C.Roland$^4$,
G.Roland$^4$,
L.Rosenberg$^4$,
J.Sagerer$^6$,
P.Sarin$^4$,
P.Sawicki$^3$,
W.Skulski$^8$,
S.G.Steadman$^4$,
P.Steinberg$^2$,
G.S.F.Stephans$^4$,
M.Stodulski$^3$,
A.Sukhanov$^2$,
J.-L.Tang$^5$,
R.Teng$^8$,
A.Trzupek$^3$,
C.Vale$^4$,
G.J.van~Nieuwenhuizen$^4$,
R.Verdier$^4$,
B.Wadsworth$^4$,
F.L.H.Wolfs$^8$,
B.Wosiek$^3$,
K.Wo\'{z}niak$^3$,
A.H.Wuosmaa$^1$,
B.Wys\l ouch$^4$\\
}
\mbox{}\\ 
{\footnotesize 
$^1$ Argonne National Laboratory, 
$^2$ Brookhaven National Laboratory, 
$^3$ Institute of Nuclear Physics, Krak\'{o}w, Poland, 
$^4$ Massachusetts Institute of Technology,
$^5$ National Central University, Chung-Li, Taiwan, 
$^6$ University of Illinois at Chicago, 
$^7$ University of Maryland,
$^8$ University of Rochester}
}

\begin{document}

\maketitle

\begin{abstract}
The PHOBOS experiment at RHIC has recorded measurements for Au-Au
collisions spanning nucleon-nucleon center-of-mass energies from
$\sqrt{s_{_{NN}}} =$ 19.6 GeV to 200 GeV.
Global observables such as elliptic flow and charged particle 
multiplicity provide important constraints on model predictions that
characterize the state of matter produced in these collisions.  
The nearly 4$\pi$ acceptance of the PHOBOS experiment provides excellent 
coverage for complete flow and multiplicity measurements.
Results including beam energy and centrality dependencies are
presented and compared to elementary systems.
\end{abstract}

\section{Introduction}


The results presented here are based on PHOBOS data taken during 
the first two RHIC physics runs for Au-Au collisions at
$\sqrt{s_{_{NN}}} =$ 19.6, 130 and 200 GeV.
Event centrality (impact parameter) is characterized
by the number of participating nucleons, $\np$, 
allowing direct comparison to elementary systems, like 
$\pp$, $\pbarp$ and $\epem\rightarrow{\rm hadrons}$.
The PHOBOS detector consists mainly of silicon pad sensors to 
perform particle tracking, vertex detection and multiplicity 
measurements. 
This set of detectors has nearly full azimuthal coverage over
a large pseudorapidity range $|\eta| <$~5.4.
Details of the layout of
the Si sensors can be found elsewhere \cite{phobos2,phobos3}.
Our methods for event triggering and centrality selection have 
been previously described \cite{phobosprl,baker}.
The raw data came in the form of energy depositions from the 
passage of charged particles above threshold through individual 
Si pads, known as hits. 
The position of the primary collision vertex was determined on an
event-by-event basis by extrapolating tracks found in the
spectrometer arms and/or the vertex detector.

\section{Flow}

\begin{figure}[t]
\begin{minipage}[t]{80mm}
\centerline{ \epsfig{file=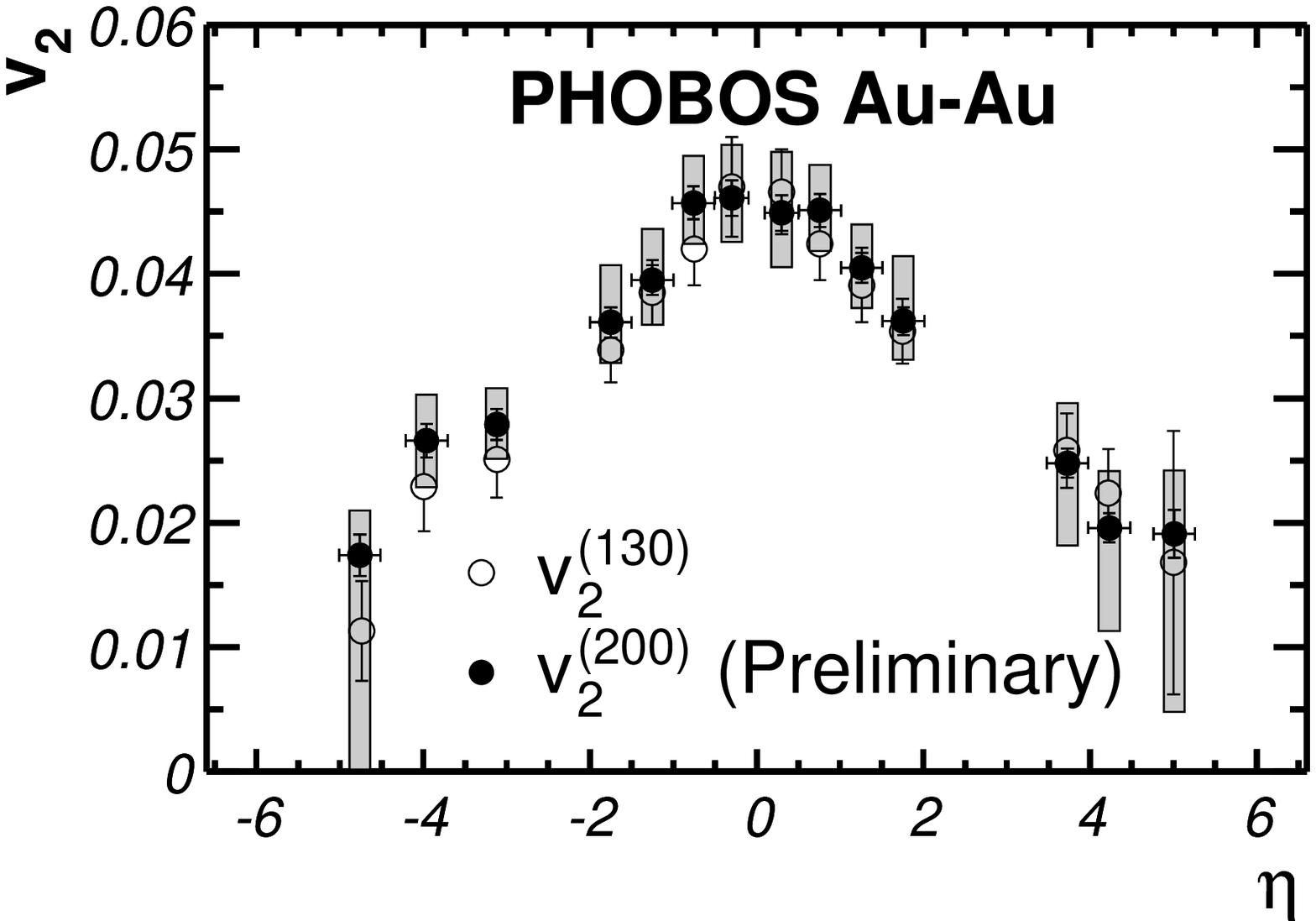,width=8cm} }
\caption{Elliptic flow as a function of pseudorapidity
for Au-Au collisions at $\sqrt{s_{_{NN}}}$ = 130 and 200 GeV.
Error bars are statistical, and boxes represent 90\% 
confidence level systematic errors on the 200 GeV data. 
}
\label{v2etaener}
\end{minipage}
\hspace{\fill}
\begin{minipage}[t]{75mm}
\centerline{ \epsfig{file=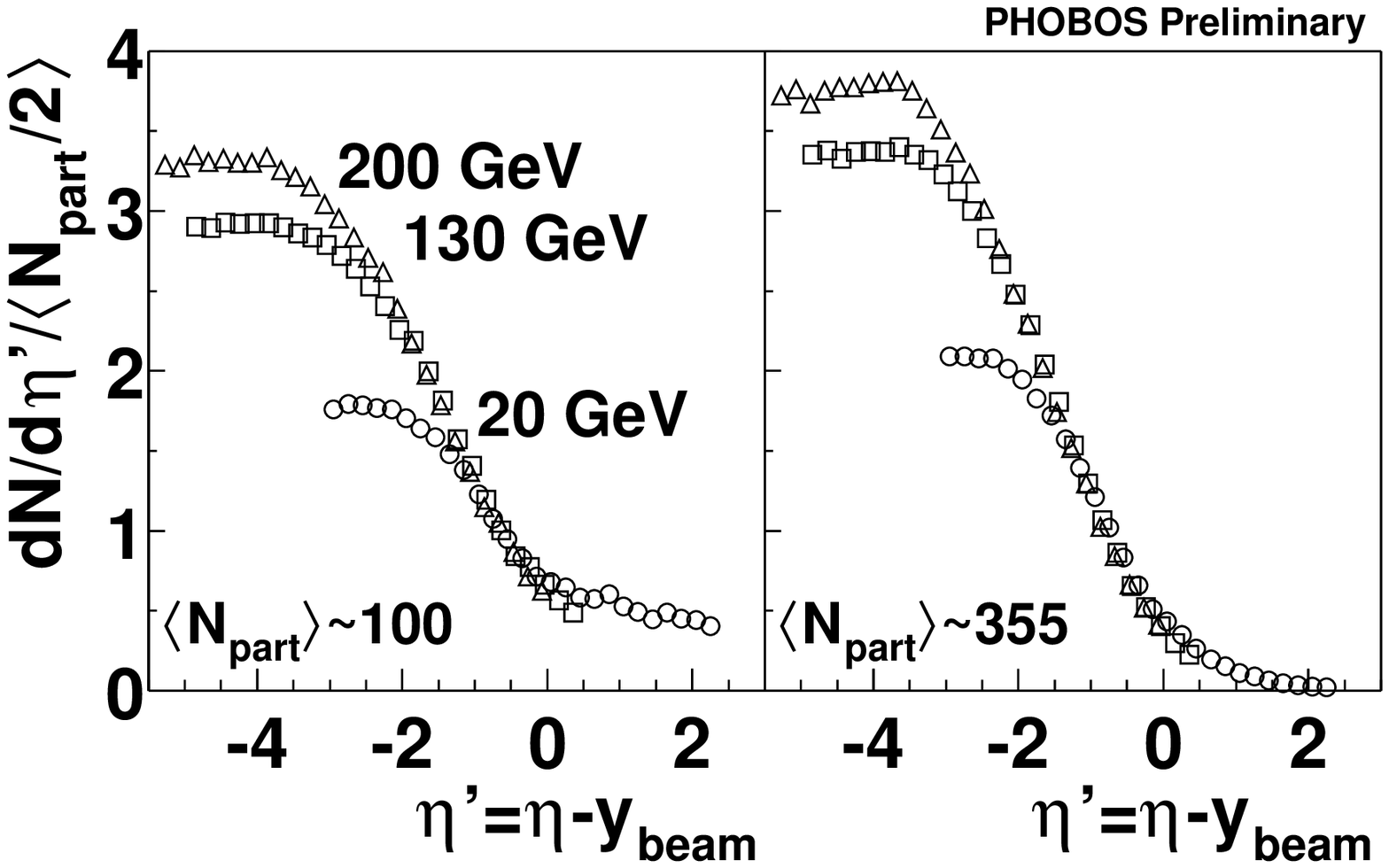,width=8.5cm} }
\caption{Charged particle pseudorapidity distributions scaled 
by the number of participant pairs for central (non-central)
Au-Au collisions at $\sqrt{s_{_{NN}}}$ = 19.6, 130 and 200 GeV
are shown in the right (left) panel.  Systematic errors are 
not shown.}
\label{limfrag}
\end{minipage}
\end{figure}

The particle emission pattern 
created in heavy-ion collisions can be probed by measuring
azimuthal anisotropy about the event plane, {\it i.e.}, elliptic flow. 
The event plane was determined by a standard subevent technique
\cite{pandv} using hits in symmetric and uniform regions of the
octagonal barrel multiplicity detector.
The second Fourier coefficient of the hit azimuthal angle distribution, 
v$_{2}$, (elliptic flow) was evaluated by correlating
the event plane to hits in a different region of the multiplicity
detectors and corrected for event plane resolution.
Details of this flow analysis are provided in Ref.~\cite{flow130}.

Flow results are shown in Fig.~\ref{v2etaener},
where the error bars indicate 1$\sigma$ statistical errors.
The 90\% confidence level systematic errors are shown as boxes 
for the 200 GeV data points. 
The flow signal changes little with the increase in the center-of-mass
energy of the collision from 130 to 200 GeV.  
Fig.~\ref{v2etaener} shows a substantial drop in v$_{2}$ as a function 
of $|\eta|$ in contradiction with the idea of longitudinal boost 
invariance over a broad rapidity range in RHIC collisions.


\begin{figure}[htb]
\begin{minipage}[t]{80mm}
\centerline{ \epsfig{file=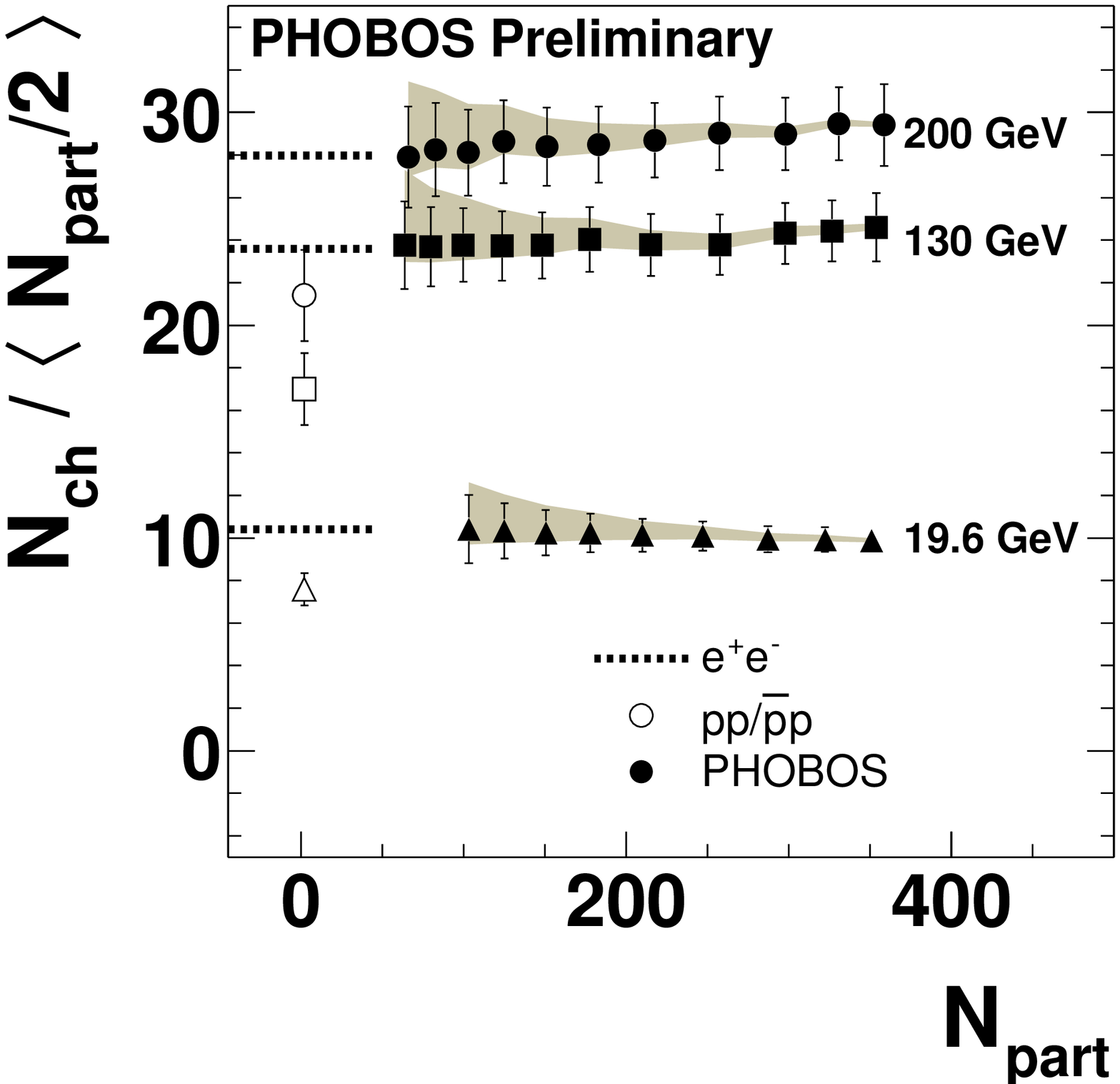,width=8cm} }
\caption{Centrality dependence of the total charged 
particle multiplicity per participant pair in Au-Au collisions 
at $\sqrt{s_{_{NN}}}$ = 19.6, 130 and 200 GeV. 
Error bars include systematic error.  
The shaded band shows the uncertainty from high-$\eta$ extrapolation.
Also shown are results for $\epem$ and $\pbarp$ data.} 
\label{ntot}
\end{minipage}
\hspace{\fill}
\begin{minipage}[t]{75mm}
\centerline{ \epsfig{file=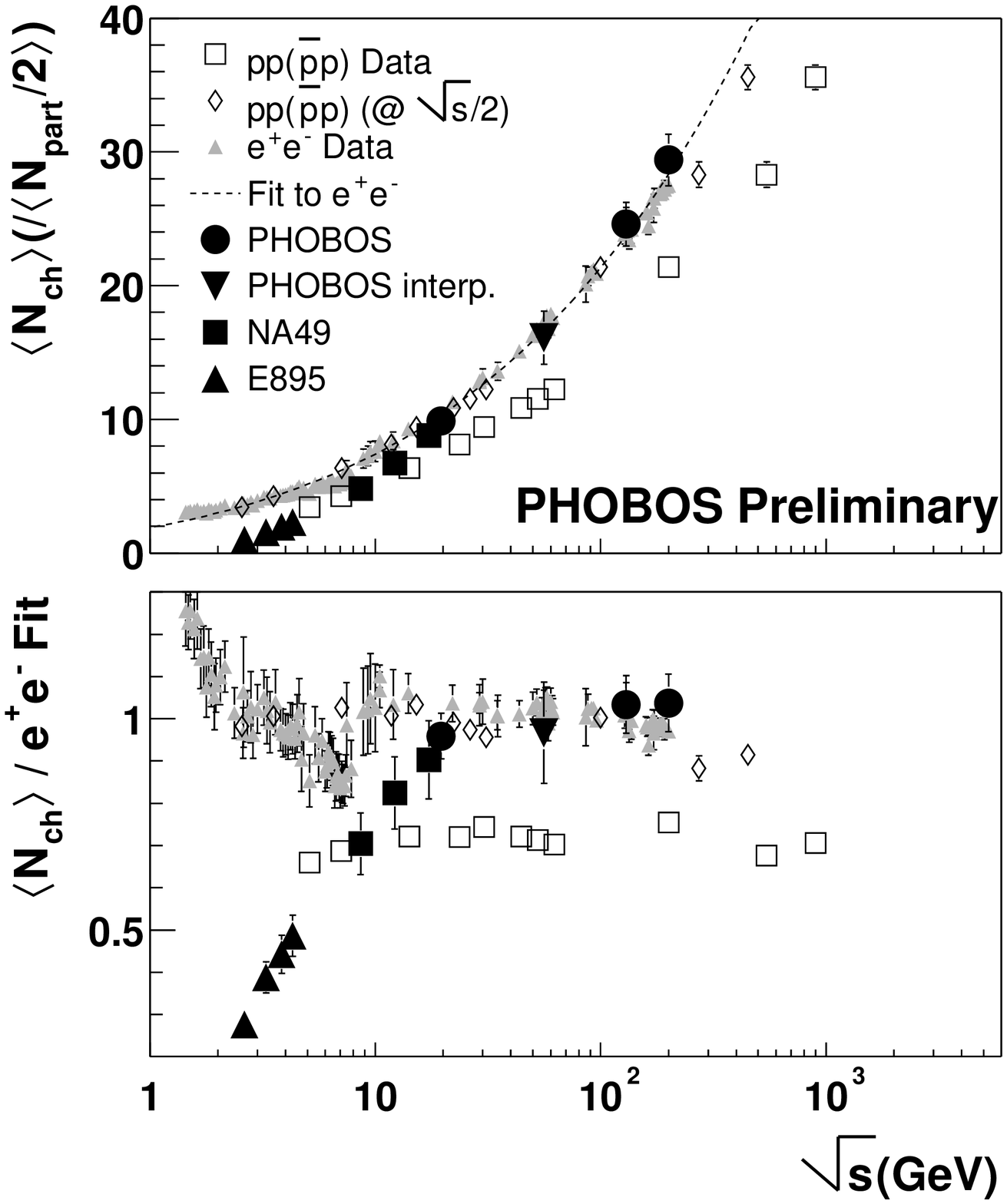,width=8cm} }
\caption{Comparison of the total charged particle multiplicity for 
elementary and heavy-ion collisions as a function of collision 
energy, as described in the text.
In the bottom panel the data are divided by the $e^+e^-$ fit shown 
in the top panel.} 
\label{total_ratio}
\end{minipage}
\end{figure}

\section{Multiplicity}

Fig.~\ref{limfrag} shows charged particle pseudorapidity
distributions scaled by the number of participant pairs, 
$\dndetapnp$ (where $\etap = \eta-\yb$), 
measured at three different
RHIC energies for peripheral ($\avenp \sim 100$) and central events 
($\avenp \sim 355$),
in the left and right panels, respectively.
Shifting into the rest frame of one
of the colliding nuclei clearly demonstrates the ``limiting
behavior'' in the fragmentation region,
{\it i.e.}, the distributions are independent of beam energy 
over a substantial range in $\etap$ \cite{baker}.
Similar behavior has been observed in $\pbarp$ \cite{ua5},
$pA$ \cite{pA}, and $\epem$ \cite{delphi} collisions over a 
large range of energies.
In  Fig.~\ref{limfrag}
the extent of the limiting fragmentation region significantly grows
with collision energy in sharp contrast to a boost-invariant scenario. 

Comparison of the data at a fixed beam energy between the left and 
right panels of Fig.~\ref{limfrag}
demonstrates that even though the shapes of the distributions 
dramatically change, the extent of the fragmentation region is nearly 
independent of centrality.
The total charged particle multiplicity, $\nch$, was determined by
integrating $\dndetap$, where the low energy data 
along with fits to the higher energy data
was used to extrapolate out to larger $\etap$ for the higher energies.
Repeating this procedure at each beam energy for all centrality 
classes results in Fig.~\ref{ntot}, which shows $\ratio$ is 
centrality independent.


In Fig.~\ref{total_ratio}, $\ratio$ in central heavy-ion 
collisions \cite{heavy-ion} is compared 
to $\epem$ and $\pp/\pbarp$ data over a large range in 
collision energy, $\s$ \cite{Groom:in}.
$\ratio$ is observed to lie below $\pp$ at low energies,
passing through the $\pp$ data around $\s\sim 10$ GeV, and then 
gradually converging with the $\epem$ trend above CERN SPS energies.
These comparisons can be made clearer by dividing all of the
data with a fit to the $\epem$ data \cite{Mueller:cq} as shown in
the lower panel of Fig.~\ref{total_ratio}.
The $\pp/\pbarp$ data (open squares) follows the same trend 
as $\epem$, but agrees better if rescaled to an ``effective energy'' 
$\seff = \s/2$ (open diamonds), which 
approximately accounts for the leading particle effect seen
in $\pp$ collisions \cite{basile}.  
Ref.~\cite{basile} finds that bulk particle
production in $\pp$ and $\epem$ data
does not depend in detail on the collision system, but 
rather the energy available for particle production.
In this scenario, the Au-Au data suggests a substantially reduced 
leading particle effect in central collisions of heavy nuclei at 
high energy \cite{peter}.

Modulation of the leading particle effect 
in nuclear collisions may be the result of multiple scattering.
For example, each participating nucleon is typically struck 
three times on average as it passes through the oncoming 
Au nucleus for $\np>65$.  
This rescattering could transfer much more of the 
initial longitudinal energy into particle production.
This naturally leads to the scaling of total particle production
in heavy-ion collisions with $\np$, as seen in Fig.~\ref{ntot},
suggestive of the ``wounded nucleon model'' \cite{wounded}
but with the scaling factor determined by $\epem$ rather 
than $\pp$ \cite{peter}.

\section{Summary}

Recent results from PHOBOS at RHIC indicate clear evidence of 
limiting fragmentation in Au-Au collisions.
Also, longitudinal boost invariance occurs only over a limited
region, because this fragmentation region grows with beam energy,
along with the pronounced drop in flow at high $\eta$.
Finally, there is a striking similarity of particle 
production in $\epem$ and particle production per 
participating nucleon pair in Au-Au collisions.



\mbox{  }

{\footnotesize {\bf Acknowledgments:}  This work was partially
supported by US DoE grants DE-AC02-98CH10886, DE-FG02-93ER40802,
DE-FC02-94ER40818, DE-FG02-94ER40865, DE-FG02-99ER41099,
W-31-109-ENG-38, and NSF grants 9603486, 9722606 and 0072204. The
Polish group was partially supported by KBN grant 2-P03B-10323.
The NCU group was partially supported by NSC of Taiwan under
contract NSC 89-2112-M-008-024.}


\begin{thebibliography}{9}
\bibitem{phobos2} H.~Pernegger {\it et al.},
Nucl. Inst. Meth. {\bf A473} (2001) 197.
\bibitem{phobos3} B.~B.~Back {\it et al.},
Nucl. Phys {\bf A698} (2002) 416c.
\bibitem{phobosprl} B.~B.~Back {\it et al.}, 
Phys. Rev. C {\bf 65} (2002) 061901R.
\bibitem{baker} B.~B.~Back {\it et al.},
arXiv:nucl-ex/0210015, submitted to Phys.\ Rev.\ Lett.\
\bibitem{pandv}  A.~M.~Poskanzer and S.~A.~Voloshin, Phys. Rev. C 
{\bf 58} (1998) 1671.
\bibitem{flow130} B.~B.~Back {\it et al.},
Phys. Rev. Lett. {\bf 89} (2002) 222301.
\bibitem{ua5} G.~J.~Alner {\it et al.}, Z.\ Phys.\ C {\bf 33} 
(1986) 1.
\bibitem{pA} J.~E.~Elias {\it et al.}, Phys. Rev. D {\bf 22}
(1980) 13.
\bibitem{delphi} P.~Abreu {\it et al.}, Phys.\ Lett.\ B {\bf 459} 
(1999) 397.
\bibitem{heavy-ion} J. Klay, U.C. Davis PhD. Thesis (2001),  
S.~V.~Afanasiev {\it et al.}, arXiv:nucl-ex/0205002.
\bibitem{Groom:in} D.~E.~Groom {\it et al.}, Eur.\ Phys.\ J.\ C 
{\bf 15} (2000) 1.
\bibitem{Mueller:cq} A.~H.~Mueller, Nucl.\ Phys.\ B {\bf 213}
(1983) 85.
\bibitem{basile} M.~Basile {\it et al.}, Phys.\ Lett.\ B {\bf 92}
(1980) 367; Phys.\ Lett.\ B {\bf 95} (1980) 311.
\bibitem{peter} P.~Steinberg {\it et al.}, arXiv:nucl-ex/0210024. 
\bibitem{wounded} J.~E.~Elias {\it et al.} Phys.\ Rev.\ Lett.\  
{\bf 41} (1978) 285, A.~Bia\l as, B.~Bleszy\'{n}ski and 
W.~Czy\.{z}, Nucl.\ Phys.\ {\bf B111} (1976) 461.
\end{thebibliography}
\end{document}